\documentclass[prd,nofootinbib,english]{revtex4}
\usepackage{amsmath}
\usepackage{amsfonts,color}
\usepackage{amsmath}
\usepackage{amsthm}
\usepackage{pdfpages}
\usepackage{amsfonts,color}
\usepackage{amssymb,float}

\usepackage[utf8]{inputenc}
\allowdisplaybreaks
\usepackage{color}
\usepackage{hyperref}
\hypersetup{
    colorlinks=true,
    linkcolor=blue,
    filecolor=magenta,      
   citecolor=blue
}

\usepackage{enumitem}

\usepackage{tikz}
\usetikzlibrary{shapes.geometric}
\usetikzlibrary{arrows.meta,arrows}

\begin{document}
\title{New models with independent dynamical torsion and nonmetricity fields}

\author{Sebastian Bahamonde$^{1,2,}$}
\email{sbahamonde@ut.ee, sebastian.beltran.14@ucl.ac.uk}
\author{Jorge Gigante Valcarcel$^{1,}$}
\email{jorge.gigante.valcarcel@ut.ee}
\affiliation{$^{1}$Laboratory of Theoretical Physics, Institute of Physics, University of Tartu, W. Ostwaldi 1, 50411 Tartu, Estonia \\
$^{2}$Laboratory for Theoretical Cosmology, Tomsk State University of
Control Systems and Radioelectronics, 634050 Tomsk, Russia (TUSUR)}

\begin{abstract}

We propose a gravitational model which allows the independent dynamical behaviour of the torsion and nonmetricity fields to be displayed in the framework of Metric-Affine gauge theory of gravity. For this task, we derive a new exact black hole solution referred to this model which extends the role of torsion of the main well-known exact solutions based on Weyl-Cartan geometry and constitutes the first known isolated gravitational system characterized by a metric tensor with independent spin and dilation charges.

\end{abstract}

\maketitle

\section{Introduction}

General Relativity (GR) is one of the most successful and accurate theories in Physics, describing gravity as a geometrical property of the space-time. The gravitational interaction is then characterized by a metric tensor and a symmetric metric-compatible affine connection, which are successfully described under the projective invariance of the Einstein-Hilbert action \cite{Dadhich:2010xa,Afonso:2017bxr}. In this standard formulation, the metric tensor turns out to be the only dynamical variable and the gravitational field is fully ascribed to the curvature tensor of Riemannian geometry. Nevertheless, the introduction of a general affine connection composed by additional pieces, namely the torsion and nonmetricity tensors, can be appropriately considered to provide equivalent formulations of the gravitational phenomena in the framework of post-Riemannian geometry \cite{BeltranJimenez:2019tjy}.

Despite the success of these representations, there are still many theoretical and observational questions such as the nature of the dark components of the universe, the renormalization of the quantum regime or the problem of singularities. One possible route to try to tackle these issues is to extend the role of the geometric structure present in the theoretical framework towards a full post-Riemannian description of gravity with dynamical metric, torsion and nonmetricity tensors. Such an extension can be related to the existence of a new fundamental symmetry in nature by applying the gauge principles not only to the external rotations and translations but also to the scale and shear transformations, which leads to the formulation of the Metric-Affine Gauge theory of gravity (MAG) \cite{Hehl:1994ue}.

Accordingly, a gauge invariant Lagrangian can be constructed from the generalized field strength tensors of this framework to introduce the dynamical aspects of the gravitational field, in such a way that the propagating role of torsion and nonmetricity can be only provided by higher order corrections in the curvature tensor, even in the absence of matter fields. The distinct restrictions on the Lagrangian coefficients lead to a large class of gravitational models where an extensive number of fundamental differences may arise. This fact evinces that the search and analysis of exact solutions are essential to improve the understanding and the physical implications of the dynamics provided by MAG. In this sense, the well-known classes of exact black hole solutions present in MAG are characterized by nontrivial torsion and nonmetricity tensors constrained by common restrictions, such as the so-called triplet ansatz or other similar assumptions obtained by prolongation techniques \cite{Tresguerres:1995js,tresguerres1995exact,Tucker:1995fw,Obukhov:1996pf,Vlachynsky:1996zh,Obukhov:1996ka,ho1997some,Macias:1998fe,Garcia:1998jw,Hehl:1999sb,Heinicke:2005bp,Baekler:2006de}. Thereby, the resulting geometry induced by these fields only embodies a common feature and does not show a full dynamical correspondence between the metric, the coframe and the corresponding independent post-Riemannian structures present in the connection.

In the present work, we apply an extension of the gravitational models with dynamical torsion recently proposed in \cite{Cembranos:2016gdt,Cembranos:2017pcs}, in order to obtain a nontrivial configuration with independent dynamical torsion and nonmetricity fields in the realm of the Weyl-Cartan geometry provided by MAG. For this task, we organise this paper as follows. In Sec.~\ref{sec:metricaffine}, we give a brief overview about MAG and the general features provided by an independent affine connection. In Sec.~\ref{sec:action}, we define a viable gravitational action with higher order corrections containing both dynamical torsion and nonmetricity fields. In addition, we also derive the field equations for this model and show that it encompasses its proper weak-field limit in the torsion and nonmetricity tensors. Sec.~\ref{sec:spherical} is devoted to the study of the static and spherically symmetric case. Then we show how the dynamical contributions of torsion and nonmetricity can be decomposed in a regular way under appropriate consistency constraints. Such a decomposition certainly simplifies the complexity of the highly nonlinear character of the field equations and allows us to solve these equations in a systematic way. The resulting exact black hole solution shows a Reissner-Nordstr\"{o}m-like geometry and carries both spin and dilation charges. Finally, we present the conclusions of our work in Sec.~\ref{sec:conclusions}.

We work in the natural units $c=G=1$ with the metric signature $(+,-,-,-)$. Quantities denoted with a tilde on top denote that they are computed with respect to a generic connection whereas quantities without a tilde denote that they are computed with respect to the Levi-Civita connection (see Table~\ref{tab:notation} for a detail list of our notational conventions). Finally, Latin and Greek indices refer to anholonomic and coordinate basis, respectively.

\section{Metric-Affine gravity with torsion and nonmetricity }\label{sec:metricaffine}

From a geometrical point of view, the standard framework of GR can be consistently formulated as a particular case of a more general class of metric-affine theories, where the geometry of the space-time is described by a metric, a coframe and an independent linear connection \cite{Hehl:1994ue}. Accordingly, the affine connection encodes additional post-Riemannian degrees of freedom, which indeed represent the torsion and nonmetricity deformations of an affinely connected metric space-time:
\begin{equation}
T^{\lambda}\,_{\mu \nu}=2\tilde{\Gamma}^{\lambda}\,_{[\mu \nu]}\,,
\end{equation}
\begin{equation}
Q_{\lambda \mu \nu}=\tilde{\nabla}_{\lambda}g_{\mu \nu}\,.
\end{equation}
The components of the affine connection can then be split into independent pieces as follows:
\begin{equation}
\tilde{\Gamma}^{\lambda}\,_{\mu \nu}=\Gamma^{\lambda}\,_{\mu \nu}+K^{\lambda}\,_{\mu \nu}+L^{\lambda}\,_{\mu \nu}\,,
\end{equation}
where $K^{\lambda}\,_{\mu \nu}$ is a metric-compatible contortion tensor containing torsion and $L^{\lambda}\,_{\mu \nu}$ a disformation tensor depending on nonmetricity:
\begin{equation}
K^{\lambda}\,_{\mu \nu}=\frac{1}{2}\left(T^{\lambda}\,_{\mu \nu}-T_{\mu}\,^{\lambda}\,_{\nu}-T_{\nu}\,^{\lambda}\,_{\mu}\right)\,,
\end{equation}
\begin{equation}
L^{\lambda}\,_{\mu \nu}=\frac{1}{2}\left(Q^{\lambda}\,_{\mu \nu}-Q_{\mu}\,^{\lambda}\,_{\nu}-Q_{\nu}\,^{\lambda}\,_{\mu}\right)\,.
\end{equation}
The resulting geometric structure is then provided by a metric tensor and an asymmetric affine connection which in general does not preserve the lengths and angles of vectors under parallel transport. Thus, such a connection introduces modifications in the covariant derivative which indeed involves a change on its commutation relations when considering an arbitrary vector $v^{\lambda}$:
\begin{equation}
[\tilde{\nabla}_{\mu},\tilde{\nabla}_{\nu}]\,v^{\lambda}=\tilde{R}^{\lambda}\,_{\rho \mu \nu}\,v^{\rho}+T^{\rho}\,_{\mu \nu}\,\tilde{\nabla}_{\rho}v^{\lambda}\,,
\end{equation}
where the corresponding curvature tensor acquires the following form:
\begin{equation}\label{totalcurvature}
\tilde{R}^{\lambda}\,_{\rho \mu \nu}=\partial_{\mu}\tilde{\Gamma}^{\lambda}\,_{\rho \nu}-\partial_{\nu}\tilde{\Gamma}^{\lambda}\,_{\rho \mu}+\tilde{\Gamma}^{\lambda}\,_{\sigma \mu}\tilde{\Gamma}^{\sigma}\,_{\rho \nu}-\tilde{\Gamma}^{\lambda}\,_{\sigma \nu}\tilde{\Gamma}^{\sigma}\,_{\rho \mu}\,.
\end{equation}

Furthermore, the inclusion of nonmetricity into the geometrical framework allows the definition of two different traces of the curvature tensor, namely the Ricci and co-Ricci tensors\footnote{Note that the trace of the Ricci and co-Ricci tensors provides a unique independent scalar curvature.}:
\begin{eqnarray}\label{Riccitensor}
\tilde{R}_{\mu\nu}&=&\tilde{R}^{\lambda}\,_{\mu \lambda \nu}\,,\\
\label{co-Riccitensor}
\hat{R}_{\mu\nu}&=&\tilde{R}_{\mu}\,^{\lambda}\,_{\nu\lambda}\,,
\end{eqnarray}
whereas a homothetic component $\tilde{R}^{\lambda}\,_{\lambda\mu\nu}$ depending on the trace part of nonmetricity can be also defined to encode the change of lengths provided by this quantity.

A gauge approach to gravity arises naturally when the unitary irreducible representations of relativistic particles labeled by their spin and mass are linked to the geometry of the space-time. Then, a gauge connection of the Poincar\'{e} group $ISO(1,3)$ can be introduced to describe the gravitational field as a gauge field of the external rotations and translations \cite{Hehl:1976kj}. Thereby not only an energy-momentum
tensor of matter arises from this approach, but also a nontrivial spin density tensor which operates as source of torsion, providing an extended correspondence between the geometry of the space-time and the properties of matter.

Nevertheless, such a connection only represents the metric-compatible part of a general linear connection, which means that the gauge formalism can be generalized in terms of the affine group defined by the semidirect product of the translation group $R^{4}$ and the general linear group $GL(4,R)$, in order to include additional intrinsic properties of matter into the geometrical scheme \cite{Hehl:1994ue,Blagojevic:2013xpa}. In this case, the resulting hypermomentum current also includes dilation and shear currents related to the trace and traceless parts of the nonmetricity tensor, respectively \cite{McCrea:1992wa}:
\begin{equation}
Q_{\lambda\mu\nu}=g_{\mu\nu}W_{\lambda}+{\nearrow\!\!\!\!\!\!\!Q}_{\lambda\mu\nu}\,,
\end{equation}
where $W_{\lambda}$ is the so-called Weyl vector. The existence of a general metalinear group $\overline{GL}(4,R)$ as a double covering of $GL(4,R)$ enables the description of fermionic matter coupled to both torsion and nonmetricity tensors in terms of infinite-dimensional spinor representations, which in the absence of nonmetricity are accordingly reduced to an infinite direct sum of spinors related to the Poincar\'{e} subgroup \cite{Neeman:1977iup,Neeman:1987pzd}. In any case, in the framework of Weyl-Cartan geometry the nonmetricity tensor is fully determined by the Weyl vector and the group $GL(4,R)$ is reduced to the homogeneous Weyl group, so that the spinor representations can be straightforwardly induced from the double covering of the Lorentz group, in virtue of the commutation between such a covering group and the scale group.

By taking into account the usual relations of the generators $P_{a}$ and $L_{a}\,^{b}$ of the affine group:
\begin{eqnarray}
\left[P_{a},P_{b}\right]&=&0\,,
\\
\left[L_{a}\,^{b},P_{c}\right]&=&i\,\delta^{b}\,_{c}\,P_{a}\,,
\\
\left[L_{a}\,^{b},L_{c}\,^{d}\right]&=&i\left(\delta^{b}\,_{c}\,L_{a}\,^{d}-\delta_{a}\,^{d}\,L_{c}\,^{b}\right)\,,
\end{eqnarray}
it is possible to obtain the following gauge curvatures from the anholonomic metric, coframe and connection:
\begin{eqnarray}
G_{ab\mu}&=&\partial_{\mu}g_{ab}-g_{ac}\,\omega^{c}\,_{b\mu}-g_{bc}\,\omega^{c}\,_{a\mu}\,,
\\
F^{a}\,_{\mu\nu}&=&\partial_{\mu}e^{a}\,_{\nu}-\partial_{\nu}e^{a}
\,_{\mu}+\omega^{a}\,_{b\mu}\,e^{b}\,_{\nu}-\omega^{a}\,_{b\nu}\,e^{b}\,_{\mu}\,,
\\
F^{a}\,_{b\mu\nu}&=&\partial_{\mu}\omega^{a}\,_{b\nu}
-\partial_{\nu}\omega^{a}\,_{b\mu}+\omega^{a}\,_{c\mu}
\,\omega^{c}\,_{b}\,_{\nu}-\omega^{a}\,_{c\nu}
\,\omega^{c}\,_{b\mu}\,.
\end{eqnarray}
These quantities characterize the properties of the gravitational interaction, which are potentially modified by the presence of torsion and nonmetricity in the framework of MAG. In particular, they are related to the nonmetricity, torsion and curvature tensors as follows:
\begin{eqnarray}
G_{ab\mu}&=&\,g_{a c}g_{b d}e^{c\lambda}e^{d\rho}Q_{\mu\lambda\rho},
\\
F^{a}\,_{\mu\nu}&=&e^{a}\,_{\lambda}T^{\lambda}\,_{\nu\mu}\,,
\\
F^{a}\,_{b\mu\nu}&=&g_{b c}\,e^{a}\,_{\lambda}e^{c\rho}\tilde{R}^{\lambda}\,_{\rho\mu\nu}\,.
\end{eqnarray}

The structure of the $GL(4,R)$ group then allows the definition of a large number of scalar invariants depending on the aforementioned tensors, in such a way that the introduction of higher order curvature corrections into the gravitational action generally involves a dynamical regime, even in the absence of matter fields.

\section{Gravitational action and field equations}\label{sec:action}
As mentioned above, the propagating character of torsion and nonmetricity demands the presence of higher order curvature terms in the gravitational action. By contrast, the Einstein-Hilbert Lagrangian defined by the Riemannian scalar curvature is sufficient to provide the dynamics of the gravitational aspects based on the metric curvature of GR. Hence a minimal extension of GR can be achieved to incorporate the propagation of torsion and nonmetricity by considering a MAG model in a background given by the proper framework of GR (i.e. the framework of GR is completely recovered when the dynamical character of the torsion and nonmetricity fields vanishes). In that case, we consider the following minimal extension for the curvature densities of the corresponding gravitational action to encompass such a limit:
\begin{eqnarray}
\tilde{R}_{\lambda\rho\mu\nu}\tilde{R}^{\lambda\rho\mu\nu} &\; \rightarrow \;& \frac{1}{2}\,\tilde{R}_{\lambda\rho\mu\nu}\left(\tilde{R}^{\lambda\rho\mu\nu}-\tilde{R}^{\rho\lambda\mu\nu}\right)\,,\\
\tilde{R}_{\lambda\rho\mu\nu}\tilde{R}^{\lambda\mu\rho\nu} &\; \rightarrow \;& \frac{1}{4}\,\tilde{R}_{\lambda\rho\mu\nu}\left(\tilde{R}^{\lambda\mu\rho\nu}+\tilde{R}^{\mu\rho\lambda\nu}-2\tilde{R}^{\mu\lambda\rho\nu}\right)\,,\\
\tilde{R}_{\left[\mu\nu\right]}\tilde{R}^{\left[\mu\nu\right]} &\; \rightarrow \;& \frac{1}{2}\left(\tilde{R}_{\left[\mu\nu\right]}\tilde{R}^{\left[\mu\nu\right]}+\hat{R}_{\left[\mu\nu\right]}\hat{R}^{\left[\mu\nu\right]}\right)\,.
\end{eqnarray}
Thereby, each one of these correspondences becomes an equivalence in the absence of nonmetricity and the corresponding independent invariants are appropriately recovered in the gravitational action with dynamical torsion \cite{Cembranos:2016gdt,Cembranos:2017pcs}. Furthermore, in the realm of Weyl-Cartan geometry, such a minimal extension also allows the quantities $\tilde{R}_{\rho\lambda\mu\nu}\tilde{R}^{\mu\lambda\rho\nu}+2\tilde{R}_{\lambda\rho\mu\nu}\tilde{R}^{\mu\lambda\rho\nu}+\tilde{R}_{\lambda\rho\mu\nu}\tilde{R}^{\lambda\mu\rho\nu}$ and $\tilde{R}_{\lambda\rho\mu\nu}\tilde{R}^{(\lambda\rho)\mu\nu}$ to be introduced into the gravitational scheme as nontrivial curvature corrections provided by nonmetricity. Nevertheless, the latter can be equivalently expressed in terms of the square of the homothetic curvature, as can be noticed by the Bianchi identity related to the nonmetricity tensor:
\begin{equation}\label{nonmetricitybianchi}
\tilde{R}^{\left(\lambda\rho\right)}\,_{\mu\nu}=\tilde{\nabla}_{[\nu}Q_{\mu]}\,^{\lambda\rho}+\frac{1}{2}\,T^{\sigma}\,_{\mu\nu}Q_{\sigma}\,^{\lambda\rho}\,,
\end{equation}
which is reduced to the following relation when the nonmetricity tensor is expressed in terms of the Weyl vector:
\begin{equation}
\tilde{R}^{\left(\lambda\rho\right)}\,_{\mu\nu}=g^{\lambda\rho}\nabla_{[\nu}W_{\mu]}=\frac{1}{4}\,g^{\lambda\rho}\tilde{R}^{\sigma}\,_{\sigma\mu\nu}\,.
\end{equation}

On the other hand, for the rest of torsion and nonmetricity densities, we can deal with a nontrivial quadratic correction in the axial component of torsion, in order to recover the respective limits with vanishing dynamical torsion or nonmetricity \cite{Cembranos:2017pcs,Tresguerres:1995js}. Then, the gravitational action for such a minimal extension is written as follows:
\begin{eqnarray}\label{Lagrangian}
S&=&\int d^4x \sqrt{-g}\left\{\mathcal{L}_{m}+\frac{1}{16\pi}\Bigl[-\tilde{R}+a_{1} T_{\lambda\mu\nu}T^{\lambda\mu\nu}+a_{2} T_{\lambda\mu\nu}T^{\mu\lambda\nu}+a_{3} T^{\lambda}\,_{\lambda\nu}\Bigl(T^{\mu}\,_{\mu}\,^{\nu}+Q^{\nu\mu}\,_{\mu}-Q^{\mu\nu}\,_{\mu}\Bigr)
\Bigr.
\right.
\nonumber\\
& &
\left.
\Bigl.
\;\;\;\;\;\;\;\;\;\;\;\;\;\;\;\;\;\;\;+\left(2a_{1}+a_{2}\right)T_{\lambda\mu\nu}Q^{\nu\lambda\mu}+b_{1}Q^{\nu\lambda}\,_{\lambda}Q_{\nu}\,^{\mu}\,_{\mu}+b_{2}Q_{\lambda\mu\nu}Q^{\mu\lambda\nu}+e_{1}\tilde{R}^{\lambda}\,_{\lambda\mu\nu}\tilde{R}^{\rho}\,_{\rho}\,^{\mu\nu}\;\;\;\;\;\;\;\;\;\;\;\;\;\;\;\;\;\;
\Bigr.
\right.
\nonumber\\
& &
\left.
\Bigl.
\;\;\;\;\;\;\;\;\;\;\;\;\;\;\;\;\;\;\;+\,c_{1}\tilde{R}_{\lambda\rho\mu\nu}\left(\tilde{R}^{\left[\lambda\rho\right]\mu\nu}-\tilde{R}^{\left[\mu\nu\right]\lambda\rho}\right)+c_{2}\tilde{R}_{\lambda\rho\mu\nu}\Big(\tilde{R}^{\lambda\mu\rho\nu}-\frac{1}{2}\tilde{R}^{\mu\nu\lambda\rho}\Big)
\Bigr.
\right.
\nonumber\\
& &
\left.
\Bigl.
\;\;\;\;\;\;\;\;\;\;\;\;\;\;\;\;\;\;\;+\,d_{1}\left(\tilde{R}_{\left[\mu\nu\right]}\tilde{R}^{\left[\mu\nu\right]}+\hat{R}_{\left[\mu\nu\right]}\hat{R}^{\left[\mu\nu\right]}\right)+\frac{1}{2}\tilde{R}_{\lambda\rho\mu\nu}\left(\left(4e_{2}-3c_{2}\right)\tilde{R}^{\left(\lambda\mu\right)\rho\nu}+\left(4e_{2}+c_{2}\right)\tilde{R}^{\mu\left(\lambda\rho\right)\nu}\right)\Bigr]\right\}\,,
\end{eqnarray}
where $a_{3}=8b_{1}=-2b_{2}=-1$, $a_{1}=(1/2)\left(1-a_{2}\right)$ and $\mathcal{L}_{\rm m}$ represents the  matter Lagrangian.

The geometric corrections are accordingly mediated by both torsion and nonmetricity fields, which in fact yield a nonvanishing metric curvature described in a first approximation by Einstein’s model. The corresponding independent field equations can be obtained by performing variations with respect to the gauge potentials\footnote{Note that the variation with respect to the metric gauge potential does not provide an additional independent equation in MAG \cite{Hehl:1994ue}.}, $e^{a}\,_{\mu}$ and $\omega^{a}\,_{b\mu}$, acquiring the following form in Weyl-Cartan geometry:
\begin{eqnarray}\label{field_eq1}
16\pi\theta_{\mu}\,^{\nu} &=& 2\,G_{\mu}\,^{\nu}+16\pi\tilde{\mathcal{L}}\,\delta_{\mu}\,^{\nu}+3\left(1-2a_{2}\right)\left(g_{\mu\rho}\nabla_{\lambda}T^{\left[\lambda\nu\rho\right]}+\left(K_{\lambda\rho\mu}+L_{\lambda\rho\mu}\right)T^{\left[\lambda\nu\rho\right]}\right)-4e_{1}\tilde{R}^{\lambda}\,_{\lambda\sigma\mu}\tilde{R}^{\rho}\,_{\rho}\,^{\sigma\nu}
\nonumber\\
&&+\tilde{R}_{\lambda\rho\sigma\mu}
\Bigl[
4c_{1}\left(\tilde{R}^{\left[\lambda\rho\right]\nu\sigma}-\tilde{R}^{\left[\nu\sigma\right]\lambda\rho}\right)+4e_{2}\left(\tilde{R}^{\left(\rho\nu\right)\lambda\sigma}-\tilde{R}^{\left(\lambda\sigma\right)\rho\nu}+\tilde{R}^{\left(\lambda\nu\right)\rho\sigma}-\tilde{R}^{\left(\rho\sigma\right)\lambda\nu}\right)
\Bigr.\,
\nonumber\\
\Bigl.
&&+c_{2}\left(\tilde{R}^{\left[\lambda\nu\right]\rho\sigma}+\tilde{R}^{\left[\rho\sigma\right]\lambda\nu}+\tilde{R}^{\left[\rho\nu\right]\sigma\lambda}+\tilde{R}^{\left[\sigma\lambda\right]\rho\nu}+2\tilde{R}^{\left[\sigma\nu\right]\lambda\rho}\right)
\Bigr]\nonumber\\
&&+2d_{1}\left(\tilde{R}^{\nu}\,_{\lambda\rho\mu}\tilde{R}^{\left[\lambda\rho\right]}+\tilde{R}_{\lambda}\,^{\nu}\,_{\mu\rho}\hat{R}^{\left[\lambda\rho\right]}+\tilde{R}_{\lambda\mu}\tilde{R}^{\left[\nu\lambda\right]}+\hat{R}_{\lambda\mu}\hat{R}^{\left[\nu\lambda\right]}\right)\,,
\end{eqnarray}
\begin{eqnarray}\label{field_eq2}
16\pi\bigtriangleup^{\lambda\mu\nu}&=&2d_{1}
\Bigl[
\nabla_{\rho}\left(g^{\mu\nu}\tilde{R}^{\left[\lambda\rho\right]}-g^{\lambda\nu}\hat{R}^{\left[\mu\rho\right]}+g^{\lambda\rho}\hat{R}^{\left[\mu\nu\right]}-g^{\mu\rho}\tilde{R}^{\left[\lambda\nu\right]}\right)+\Bigl(K^{\rho\mu}\,_{\rho}+L^{\rho\mu}\,_{\rho}\Bigr)\tilde{R}^{\left[\lambda\nu\right]}-\Bigl(K^{\rho\lambda}\,_{\rho}+L^{\rho\lambda}\,_{\rho}\Bigr)\hat{R}^{\left[\mu\nu\right]}
\Bigr.
\nonumber\\
\Bigl.
&&+\Bigl(K^{\nu\lambda}\,_{\rho}+L^{\nu\lambda}\,_{\rho}\Bigr)\hat{R}^{\left[\mu\rho\right]}-\Bigl(K^{\nu\mu}\,_{\rho}+L^{\nu\mu}\,_{\rho}\Bigr)\tilde{R}^{\left[\lambda\rho\right]}+\Bigl(K^{\mu}\,_{\rho}\,^{\lambda}+L^{\mu}\,_{\rho}\,^{\lambda}\Bigr)\hat{R}^{\left[\rho\nu\right]}-\Bigl(K^{\lambda}\,_{\rho}\,^{\mu}+L^{\lambda}\,_{\rho}\,^{\mu}\Bigr)\tilde{R}^{\left[\rho\nu\right]}
\Bigr]
\nonumber\\
&&+\Bigl(\nabla_{\rho}+W_{\rho}\Bigr)
\Bigl[
4c_{1}\left(\tilde{R}^{\left[\lambda\mu\right]\rho\nu}-\tilde{R}^{\left[\rho\nu\right]\lambda\mu}\right)+c_{2}\left(\tilde{R}^{\left[\mu\nu\right]\lambda\rho}-\tilde{R}^{\left[\lambda\nu\right]\mu\rho}+\tilde{R}^{\left[\lambda\rho\right]\mu\nu}-\tilde{R}^{\left[\mu\rho\right]\lambda\nu}-2\tilde{R}^{\left[\rho\nu\right]\lambda\mu}\right)
\Bigr.
\nonumber\\
\Bigl.
&&+4e_{2}\left(\tilde{R}^{\left(\mu\nu\right)\lambda\rho}-\tilde{R}^{\left(\lambda\rho\right)\mu\nu}+\tilde{R}^{\left(\nu\lambda\right)\mu\rho}-\tilde{R}^{\left(\rho\mu\right)\lambda\nu}\right)
\Bigr]
-4e_{1}g^{\lambda \mu}\nabla_{\rho}\tilde{R}_{\sigma}\,^{\sigma\rho\nu}-3\left(1-2a_{2}\right)T^{\left[\lambda\mu\nu\right]}
\nonumber\\
&&+\Bigl(K^{\lambda}\,_{\sigma\rho}+L^{\lambda}\,_{\sigma\rho}\Bigr)
\Bigl[
4c_{1}\left(\tilde{R}^{\left[\sigma\mu\right]\rho\nu}-\tilde{R}^{\left[\rho\nu\right]\sigma\mu}\right)+c_{2}\left(\tilde{R}^{\left[\mu\nu\right]\sigma\rho}-\tilde{R}^{\left[\sigma\nu\right]\mu\rho}+\tilde{R}^{\left[\sigma\rho\right]\mu\nu}-\tilde{R}^{\left[\mu\rho\right]\sigma\nu}-2\tilde{R}^{\left[\rho\nu\right]\sigma\mu}\right)
\Bigr.
\nonumber\\
\Bigl.
&&+4e_{2}\left(\tilde{R}^{\left(\mu\nu\right)\sigma\rho}-\tilde{R}^{\left(\sigma\rho\right)\mu\nu}+\tilde{R}^{\left(\nu\sigma\right)\mu\rho}-\tilde{R}^{\left(\rho\mu\right)\sigma\nu}\right)+2d_{1}g^{\mu\nu}\tilde{R}^{\left[\sigma\rho\right]}
\Bigr]
\nonumber\\
&&+\Bigl(K^{\mu}\,_{\sigma\rho}+L^{\mu}\,_{\sigma\rho}\Bigr)
\Bigl[
4c_{1}\left(\tilde{R}^{\left[\lambda\sigma\right]\rho\nu}-\tilde{R}^{\left[\rho\nu\right]\lambda\sigma}\right)+c_{2}\left(\tilde{R}^{\left[\sigma\nu\right]\lambda\rho}-\tilde{R}^{\left[\lambda\nu\right]\sigma\rho}+\tilde{R}^{\left[\rho\sigma\right]\lambda\nu}-\tilde{R}^{\left[\rho\lambda\right]\sigma\nu}-2\tilde{R}^{\left[\rho\nu\right]\lambda\sigma}\right)
\Bigr.
\nonumber\\
\Bigl.
&&+4e_{2}\left(\tilde{R}^{\left(\sigma\nu\right)\lambda\rho}-\tilde{R}^{\left(\lambda\rho\right)\sigma\nu}+\tilde{R}^{\left(\nu\lambda\right)\sigma\rho}-\tilde{R}^{\left(\sigma\rho\right)\lambda\nu}\right)-2d_{1}g^{\lambda\nu}\hat{R}^{\left[\sigma\rho\right]}
\Bigr]\,,
\end{eqnarray}
where $\tilde{\mathcal{L}}$ represents the Lagrangian density composed by the quadratic order corrections, and the quantities $\theta_{\mu}\,^{\nu}$ and $\bigtriangleup^{\lambda\mu\nu}$ describe the canonical energy-momentum and hypermomentum density tensors of matter, respectively:
\begin{equation}
\theta_{\mu}\,^{\nu}=\frac{e^{a}\,_{\mu}}{\sqrt{- g}}\frac{\delta\left(\mathcal{L}_{m}\sqrt{- g}\right)}{\delta e^{a}\,_{\nu}}\,,
\end{equation}
\begin{equation}
\bigtriangleup^{\lambda\mu\nu}=\frac{e^{a\lambda}e_{b}\,^{\mu}}{\sqrt{- g}}\frac{\delta\left(\mathcal{L}_{m}\sqrt{-g}\right)}{\delta\omega^{a}\,_{b\nu}}\,.
\end{equation}

As can be seen from the Eq~(\ref{Lagrangian}), the propagation of torsion and nonmetricity is carried out by the antisymmetrized and homothetic parts of the curvature tensor, related to the components $\tilde{R}^{\lambda}\,_{[\mu\nu\rho]}$, $\tilde{R}_{[\mu\nu]}$ and $\tilde{R}^{\lambda}\,_{\lambda\mu\nu}$. Indeed, this expression can be equivalently written in terms of the mentioned pieces as follows\footnote{In order to obtain the Expression (\ref{LagrangianIrreducible}), we use the identity $\tilde{R}=R-2\nabla_{\mu}T^{\nu \mu}\,_{\nu}+\nabla_{\mu}Q^{\mu}\,_{\nu}\,^{\nu}-\nabla_{\mu}Q^{\nu}\,_{\nu}\,^{\mu}+\frac{1}{4}T_{\lambda \mu \nu}T^{\lambda \mu \nu}+\frac{1}{2}T_{\lambda \mu \nu}T^{\mu \lambda \nu}-T^{\lambda}\,_{\lambda\nu}T^{\mu}\,_{\mu}\,^{\nu}-T^{\lambda}\,_{\lambda\nu}Q^{\nu\mu}\,_{\mu}+T^{\lambda}\,_{\lambda\nu}Q^{\mu\nu}\,_{\mu}+T_{\lambda\mu\nu}Q^{\nu\lambda\mu}-\frac{1}{8}Q^{\nu\lambda}\,_{\lambda}Q_{\nu}\,^{\mu}\,_{\mu}+\frac{1}{2}Q_{\lambda\mu\nu}Q^{\mu\lambda\nu}$, which allows us to rewrite the first terms of the Expression (\ref{Lagrangian}) as the Einstein-Hilbert Lagrangian and an additional factor which acts as a total derivative in the gravitational action.}:
\begin{eqnarray}\label{LagrangianIrreducible}
S &=& \int d^4x \sqrt{-g}
\left\{\mathcal{L}_{m}+\frac{1}{64 \pi}\Bigl[
-4R+3\left(6c_{1}+c_{2}\right)\tilde{R}_{\lambda\left[\rho\mu\nu\right]}\tilde{R}^{\lambda\left[\rho\mu\nu\right]}+9\left(2c_{1}+c_{2}\right)\tilde{R}_{\lambda\left[\rho\mu\nu\right]}\tilde{R}^{\mu\left[\lambda\nu\rho\right]}
\Bigr.
\right.
\nonumber\\
& &
\left.
\Bigl.
\;\;\;\;\;\;\;\;\;\;\;\;\;\;\;\;\;\;\;+\,8\,d_{1}\tilde{R}_{\left[\mu\nu\right]}\tilde{R}^{\left[\mu\nu\right]}+\frac{1}{4}\left(16e_{1}+4e_{2}+4d_{1}-12c_{1}-3c_{2}\right)\tilde{R}^{\lambda}\,_{\lambda\mu\nu}\tilde{R}^{\rho}\,_{\rho}\,^{\mu\nu}
\Bigr.
\right.
\nonumber\\
& &
\left.
\Bigl.
\;\;\;\;\;\;\;\;\;\;\;\;\;\;\;\;\;\;\;+\,2\left(4c_{1}+c_{2}-2d_{1}\right)\tilde{R}_{\left[\mu\nu\right]}\tilde{R}^{\lambda}\,_{\lambda}\,^{\mu\nu}+3\left(1-2a_{2}\right)T_{\left[\lambda\mu\nu\right]}T^{\left[\lambda\mu\nu\right]}\Bigr]\right\}\,,
\end{eqnarray}
where the antisymmetrized part of curvature also satisfies its proper Bianchi identity:
\begin{equation}\label{torsionbianchi}
\tilde{R}^{\lambda}\,_{[\mu \nu \rho]}+\tilde{\nabla}_{[\mu}T^{\lambda}\,_{\nu \rho]}+T^{\sigma}\,_{[\mu \nu}\,T^{\lambda}\,_{\rho] \sigma}=0\,.\end{equation}

Hence, these quantities act as field strength tensors described by the deviation from the Bianchi identities of GR in the presence of torsion and nonmetricity. Furthermore, the model also encompasses the weak-field limit for the torsion and nonmetricity tensors involving the traces of these deviations. By applying a linear approximation on the traces of the Eq.~(\ref{field_eq2}), the following relations between these geometric tensors and the matter currents are obtained,
\begin{eqnarray}
\left(4c_{1}+c_{2}+2d_{1}\right)\left(\nabla_{\rho}\tilde{R}^{\lambda}\,_{\lambda}\,^{\rho\mu}+2\left(\nabla_{\rho}\nabla_{\lambda}T^{\lambda\rho\mu}+\nabla_{\rho}\nabla^{\rho}T^{\lambda\mu}\,_{\lambda}-\nabla_{\rho}\nabla^{\mu}T^{\lambda\rho}\,_{\lambda}\right)\right)=16\pi\left(\bigtriangleup_{\lambda}\,^{\mu\lambda}-\bigtriangleup^{\mu}\,_{\lambda}\,^{\lambda}\right)\,,\end{eqnarray}
\begin{eqnarray}
\left(d_{1}+2e_{1}+2e_{2}\right)\nabla_{\rho}\tilde{R}^{\lambda}\,_{\lambda}\,^{\rho\mu}=-\,4\pi\left(\bigtriangleup_{\lambda}\,^{\mu\lambda}+\bigtriangleup^{\mu}\,_{\lambda}\,^{\lambda}\right)\,.\end{eqnarray}

The resulting matter density with intrinsic hypermomentum can be consistently related to the generalised Frenkel condition in the classical regime \cite{Obukhov:1993pt}, which allows us to express the weak-field limit of torsion and nonmetricity in the following way
\begin{eqnarray}\label{weaktorsion}
\nabla_{\rho}\nabla_{\lambda}T^{\lambda\rho}\,_{\mu}+\nabla_{\rho}\nabla^{\rho}T^{\lambda}\,_{\mu\lambda}-\nabla_{\rho}\nabla_{\mu}T^{\lambda\rho}\,_{\lambda}=0\,,\end{eqnarray}
\begin{eqnarray}
\label{weaknonmetricity}
\nabla_{\mu}\tilde{R}^{\lambda}\,_{\lambda}\,^{\mu\nu}=0\,,\end{eqnarray}
with $4c_{1}+c_{2}+2d_{1} \neq 0$ and $d_{1}+2e_{1}+2e_{2} \neq 0$. It is worthwhile to rewrite Eq.~(\ref{weaktorsion}) in terms of the irreducible parts of torsion, giving us\footnote{Note that $\varepsilon^{\lambda}{}_{\rho\mu\nu}=\sqrt{-g}g^{\lambda\sigma}\epsilon_{\sigma\rho\mu\nu}$ is the Levi-Civita tensor and $\epsilon_{\sigma\rho\mu\nu}$ the Levi-Civita symbol.}
\begin{equation}\label{irreducibleparts}
T^{\lambda}\,_{\mu \nu}=\frac{1}{3}\left(\delta^{\lambda}\,_{\nu}T_{\mu}-\delta^{\lambda}\,_{\mu}T_{\nu}\right)+\frac{1}{6}\,\varepsilon^{\lambda}\,_{\rho\mu\nu}S^{\rho}+t^{\lambda}\,_{\mu \nu}\,,
\end{equation}
since it only contains contributions from the vector and tensor parts, as shown below,
\begin{equation}\label{irreduciblelimit}
\nabla_{\mu}\nabla^{[\mu}T^{\nu]}+\frac{3}{4}\,\nabla_{\mu}\nabla_{\lambda}t^{\lambda\mu\nu}=0\,.
\end{equation}

Therefore, the proposed model is characterized by three consistent limits: the weak-field regime for the torsion and nonmetricity tensors described by the expressions above, and the proper background of GR for the Riemannian structure.

\section{Spherical symmetry in Metric-Affine gravity}\label{sec:spherical}

\subsection{Distribution of the solutions and consistency constraints}

In virtue of the highly nonlinear character of the field equations of MAG, additional symmetry constraints must be imposed to find nontrivial solutions, such as the presence of a static and spherically symmetric space-time, which can be described by the line element
\begin{equation}
    ds^2=\Psi_{1}(r)\,dt^2-\frac{dr^2}{\Psi_{2}(r)}-r^2\left(d\theta_1^2+\sin\theta_1^2d\theta_2^2\right)\,,\label{metric}
\end{equation}
where $\Psi_{1}(r)$ and $\Psi_{2}(r)$ are positive functions of the radial coordinate. The tetrad fields can then be written in the orthonormal gauge as follows
\begin{equation}
    e^a{}_\mu=\textrm{diag}\Big(\sqrt{\Psi_{1}(r)},\frac{1}{\sqrt{\Psi_{2}(r)}},r,r\sin\theta_1\Big)\,.
\end{equation}

The invariance of the curvature tensor requires then not only the same type of invariance for the metric tensor but also for the torsion and nonmetricity tensors (i.e. the Lie derivative of torsion and nonmetricity in the direction of the Killing vector ${\xi}$ fulfills the condition $\mathcal{L}_{\xi}T^{\lambda}{}_{\mu\nu}=\mathcal{L}_{\xi}Q^{\lambda}{}_{\mu\nu}=0$), which provides the following structure in a spherically symmetric space-time~\cite{Hohmann:2019fvf}:
\begin{eqnarray}
T^{t}\,_{t r}&=&a(r) \,,\\
T^{r}\,_{t r}&=&b(r)\,,\\
T^{\theta_{k}}\,_{t \theta_{k}}&=&c(r)\,,\\
T^{\theta_{k}}\,_{r \theta_{k}}&=&g(r) \,,\\
T^{\theta_{k}}\,_{t \theta_{l}}&=&e^{a \theta_{k}}\,e^{b}\,_{\theta_{l}}\,\epsilon_{a b}\, d (r) \,,\\
T^{\theta_{k}}\,_{r \theta_{l}}&=&e^{a \theta_{k}}\,e^{b}\,_{\theta_{l}}\,\epsilon_{a b}\, h (r) \,,\\
T^{t}\,_{\theta_{k} \theta_{l}}&=&\epsilon_{k l} \, k (r)\,\sin\theta_1 \,,\\
T^{r}\,_{\theta_{k} \theta_{l}}&=&\epsilon_{k l} \, l (r)\,\sin\theta_1 \,,\\
W_\lambda&=&\left(w_{1}(r),w_{2}(r),0,0\right)\,,
\end{eqnarray}
where $\epsilon_{kl}$ is the Levi-Civita symbol in two dimensions. Thus, we must deal with eight arbitrary functions $a,b,c,g,d,h,k$ and $l$ of the radial coordinate for torsion and two additional functions $w_{1}$ and $w_{2}$ depending on the same coordinate for nonmetricity. This means that the problem of solving the field equations (\ref{field_eq1}) and (\ref{field_eq2}) turns out to be still very complicated and additional restrictions are required for this task. Furthermore, it is a well-known fact that nontrivial solutions with free functions (i.e. not totally determined by the variational equations) may arise in some MAG models \cite{ho1997some,Lenzen:1986hb,chen1994poincare,Zhytnikov:1991pu}. In the present case, the distribution of the solutions can be split into propagating or non-propagating classes. The latter are characterized by pure gauge configurations with vanishing or degenerate field strength tensors, which indeed reduce the general structure of the Eq.~(\ref{LagrangianIrreducible}). This allows us to express the nontrivial antisymmetrized part of the curvature tensor as a general sum of a degenerate homothetic curvature and an independent dynamical component:
\begin{equation}\label{curvaturedecomposition}
\tilde{R}^{\lambda}\,_{[\mu \nu \rho]}=\mathring{R}^{\lambda}\,_{[\mu \nu \rho]}+\bar{R}^{\lambda}\,_{[\mu \nu \rho]}\,,
\end{equation}
where $\mathring{R}^{\lambda}\,_{[\mu \nu \rho]}=\alpha\tilde{R}^{\sigma}\,_{\sigma[\mu\nu}\delta_{\rho]}\,^{\lambda}$ and $\alpha$ is a constant. Thereby, we expect to obtain dynamical contributions to the field equations for nontrivial expressions of these parts. 
Following the discussion above, let us consider the subsequent decomposition for the irreducible parts of the torsion tensor:
\begin{eqnarray}\label{decomposition2}
T_{\mu}&=&\mathring{T}_{\mu}+\bar{T}_{\mu}\,,
\nonumber\\
S_{\mu}&=&\mathring{S}_{\mu}+\bar{S}_{\mu}\,,
\nonumber\\
t_{\lambda\mu\nu}&=&\mathring{t}_{\lambda\mu\nu}+\bar{t}_{\lambda\mu\nu}\,,
\end{eqnarray}
with $\mathring{t}^{\lambda}\,_{\mu\lambda}=\bar{t}^{\lambda}\,_{\mu\lambda}=\varepsilon_{\sigma\lambda\mu\nu}\mathring{t}^{\lambda\mu\nu}=\varepsilon_{\sigma\lambda\mu\nu}\bar{t}^{\lambda\mu\nu}=0$, in such a way that the quantities denoted with a circle describe non-propagating modes with a vanishing field strength tensor and/or a redundant homothetic component for $\alpha \neq 0$, whereas the rest of the quantities with a bar may provide dynamical effects in the interaction.

In general, even the class of solutions associated with these non-propagating modes can by characterized by a singular behaviour. In this sense, it is possible to bypass this class of singular non-propagating solutions by analysing the torsion tensor referred to the rotated basis $\vartheta^{a}=\Lambda^{a}\,_{b}e^{b}$ given by the following vector fields

\begin{eqnarray}
\vartheta^{\hat{t}}&=&\frac{1}{2}{\Big(\frac{\Psi_1}{ \Psi_2}\Big)^{1/4}}\left\{ \left[\sqrt{\Psi_{1}\Psi_{2}}+1\right]\,dt+\left[1-\frac{1}{\sqrt{\Psi_{1}\Psi_{2}}}\right]\,dr \right\}\,,\\
\vartheta^{\hat{r}}&=&\frac{1}{2}{\Big(\frac{\Psi_1}{ \Psi_2}\Big)^{1/4}}\left\{ \left[\sqrt{\Psi_{1}\Psi_{2}}-1\right]\,dt+\left[1+\frac{1}{\sqrt{\Psi_{1}\Psi_{2}}}\right]\,dr \right\}\,,\\
\vartheta^{\hat{\theta}_{1}}&=&r\,d \theta_{1} \,,\\%
\vartheta^{\hat{\theta_{2}}}&=&r\sin\theta_{1} \, d \theta_{2} \,.
\end{eqnarray}
Note that for simplicity, we omitted  the radial dependence in the functions.
Then, we can write the gauge curvature $\mathcal{F}^{a}\,_{b c} = \vartheta^{a}\,_{\lambda}\vartheta_{b}\,^{\mu}\vartheta_{c}\,^{\nu}T^{\lambda}\,_{\nu\mu}\,$ related to the torsion tensor in this orthogonal coframe as follows,

\begin{eqnarray}
\mathcal{F}^{\hat{t}}\,_{\hat{t} \hat{r}} &=& {\frac{1}{2}\Big(\frac{\Psi_2}{\Psi_1}\Big)^{1/4}}\left\{\left[1+\sqrt{\Psi_{1}\Psi_{2}}\right]a+\left[1-\frac{1}{\sqrt{\Psi_{1}\Psi_{2}}}\right]b\right\}\,,\\
\mathcal{F}^{\hat{r}}\,_{\hat{t} \hat{r}} &=& {\frac{1}{2}\Big(\frac{\Psi_2}{\Psi_1}\Big)^{1/4}}\left\{\left[1+\frac{1}{\sqrt{\Psi_{1}\Psi_{2}}}\right]b-\left[1-\sqrt{\Psi_{1}\Psi_{2}}\right]a\right\} \,,\\
\mathcal{F}^{\hat{\theta_{1}}}\,_{\hat{t} \hat{\theta_{1}}} &=& \mathcal{F}^{\hat{\theta_{2}}}\,_{\hat{t} \hat{\theta_{2}}} = {\frac{1}{2}\Big(\frac{\Psi_2}{\Psi_1}\Big)^{1/4}}\left\{\left[1+\frac{1}{\sqrt{\Psi_{1}\Psi_{2}}}\right]c+\left[1-\sqrt{\Psi_{1}\Psi_{2}}\right]g\right\}\,,\\
\mathcal{F}^{\hat{\theta_{1}}}\,_{\hat{r} \hat{\theta_{1}}} &=& \mathcal{F}^{\hat{\theta_{2}}}\,_{\hat{r} \hat{\theta_{2}}} = {\frac{1}{2}\Big(\frac{\Psi_2}{\Psi_1}\Big)^{1/4}}\left\{\left[1+\sqrt{\Psi_{1}\Psi_{2}}\right]g-\left[1-\frac{1}{\sqrt{\Psi_{1}\Psi_{2}}}\right]c\right\} \,,\\
\mathcal{F}^{\hat{\theta_{2}}}\,_{\hat{t} \hat{\theta_{1}}} &=& - \mathcal{F}^{\hat{\theta_{1}}}\,_{\hat{t} \hat{\theta_{2}}} = {\frac{1}{2}\Big(\frac{\Psi_2}{\Psi_1}\Big)^{1/4}}\left\{\left[1+\frac{1}{\sqrt{\Psi_{1}\Psi_{2}}}\right]d+\left[1-\sqrt{\Psi_{1}\Psi_{2}}\right]h\right\} \,,\\
\mathcal{F}^{\hat{\theta_{2}}}\,_{\hat{r} \hat{\theta_{1}}} &=& - \mathcal{F}^{\hat{\theta_{1}}}\,_{\hat{r} \hat{\theta_{2}}} = {\frac{1}{2}\Big(\frac{\Psi_2}{\Psi_1}\Big)^{1/4}}\left\{\left[1+\sqrt{\Psi_{1}\Psi_{2}}\right]h-\left[1-\frac{1}{\sqrt{\Psi_{1}\Psi_{2}}}\right]d\right\} \,,\\
\mathcal{F}^{\hat{t}}\,_{\hat{\theta_{1}} \hat{\theta_{2}}} &=& {\frac{1}{2r^2}\Big(\frac{\Psi_1}{\Psi_2}\Big)^{1/4}}\left\{\left[1+\sqrt{\Psi_{1}\Psi_{2}}\right]k+\left[1-\frac{1}{\sqrt{\Psi_{1}\Psi_{2}}}\right]l\right\} \,,\\
\mathcal{F}^{\hat{r}}\,_{\hat{\theta_{1}} \hat{\theta_{2}}} &=& {\frac{1}{2r^2}\Big(\frac{\Psi_1}{\Psi_2}\Big)^{1/4}}\left\{\left[1+\frac{1}{\sqrt{\Psi_{1}\Psi_{2}}}\right]l-\left[1-\sqrt{\Psi_{1}\Psi_{2}}\right]k\right\} \,.
\end{eqnarray}

These components clearly show a potential singular behaviour in the roots of the metric functions $\Psi_{1}(r)$ and $\Psi_{2}(r)$, which may be induced in the scalar invariants constructed from torsion. Thus, we only consider a class of well-defined regular solutions (excluding the point $r=0$) provided by the following relations:
\begin{eqnarray}\label{rel1}
b(r)&=&a(r)\,\sqrt{\Psi_{1}(r)\Psi_{2}(r)}\,,\;\;\;\;\;\;c(r) = - \, g(r)\,\sqrt{\Psi_{1}(r)\Psi_{2}(r)}\,,
\;\;\;\;\;\;\;\;\;\;\;\;
\nonumber\\
d(r)&=&- \, h(r)\,\sqrt{\Psi_{1}(r)\Psi_{2}(r)}\,,\;\;\;l(r) = k(r)\,\sqrt{\Psi_{1}(r)\Psi_{2}(r)}\,.
\end{eqnarray}

Note that this regularity is also highlighted with the vanishing of the torsion invariants, as other well-known exact vacuum solutions with dynamical torsion \cite{Obukhov:2019fti}. On the other hand, the occurrence of unavoidable singularities can be established under appropriate energy conditions \cite{Cembranos:2016xqx,Cembranos:2019mcb}. 

The same reasoning can be applied for nonmetricity, obtaining the following relation for the components of the Weyl vector:
\begin{equation}\label{rel2}
w_{1}(r)=-\,w_{2}(r)\sqrt{\Psi_{1}(r)\Psi_{2}(r)}\,.
\end{equation}

A direct calculation over the weak-field limit provided by Eqs.~(\ref{weaktorsion}) and (\ref{weaknonmetricity}) allows us to obtain the solution for this vector and an additional constraint for the torsion components\footnote{The prime symbol denotes differentiation with respect to the radial coordinate.}:
\begin{eqnarray}
w_{1}(r)&=&-\,\kappa_{d}\,\int\sqrt{\frac{\Psi_{1}(r)}{\Psi_{2}(r)}}\,\frac{dr}{r^2}\,, \label{rel3}\\
b(r)&=&rc\,'(r)+c(r)+\frac{p}{r}\,
\sqrt{\frac{\Psi_{1}(r)}{\Psi_{2}(r)}}\,,\label{rel4}
\end{eqnarray}
where $\kappa_{d}$ represents the dilation charge and $p$ is an additional integration constant, which indeed can be related to the previous quantity, as shown below.

The relevance of the relations~(\ref{rel1})-(\ref{rel4}) is remarkable since they directly show the same type of decomposition (\ref{curvaturedecomposition}) for the antisymmetrized part of the curvature tensor when $p=\kappa_{d}/2$. In this case, the antisymmetrized part of the curvature tensor is expressed as the sum of a non-propagating component which indeed is proportional to the homothetic curvature and a general contribution which can propagate axial and tensor modes\footnote{Note that these expressions are obtained in the spherically symmetric case, which means that additional terms are expected to arise in more general cases (e.g. in the presence of an axisymmetric space-time).},
\begin{equation}
\tilde{R}^{\lambda}\,_{[\mu\nu\rho]}=\mathring{R}_{1}^{\lambda}\,_{[\mu \nu \rho]}+\tilde{R}_{2}^{\lambda}\,_{[\mu\nu\rho]}\,,\end{equation}
where
\begin{eqnarray}\label{curvaturedecomposition2}
\mathring{R}_{1}^{\lambda}\,_{[\mu \nu \rho]}&:=&\frac{2}{3}\delta^{\lambda}\,_{[\mu}\nabla_{\rho}\mathring{T}_{\nu]}+\nabla_{[\mu\,}\mathring{t}_{1}^{\lambda}\,_{\rho\nu]}=\frac{1}{4}\tilde{R}^{\sigma}\,_{\sigma[\mu\nu}\delta_{\rho]}\,^{\lambda}\,,\\
\tilde{R}_{2}^{\lambda}\,_{[\mu \nu \rho]}&:=&\frac{1}{6}\varepsilon^{\lambda}\,_{\sigma[\rho\nu}\nabla_{\mu]}S^{\sigma}+\nabla_{[\mu}t_{2}^{\lambda}\,_{\rho\nu]}-\frac{1}{18}\varepsilon_{\sigma\mu\nu\rho}\mathring{T}^{\lambda}S^{\sigma}+\frac{1}{4}\varepsilon^{\lambda}\,_{\omega\sigma[\rho}\mathring{t}_{1}^{\sigma}\,_{\mu\nu]}S^{\omega}
\nonumber\\
&&+\frac{1}{18}\varepsilon^{\lambda}\,_{\omega[\rho\nu}\mathring{T}_{\mu]}S^{\omega}+\frac{1}{3}\varepsilon^{\sigma}\,_{\omega[\mu\nu}\mathring{t}_{1}^{\lambda}\,_{\rho]\sigma}S^{\omega}+\mathring{t}_{1}^{\sigma}\,_{[\mu\nu}t_{2}^{\lambda}\,_{\rho]\sigma}+t_{2}^{\sigma}\,_{[\mu\nu}\mathring{t}_{1}^{\lambda}\,_{\rho]\sigma}
\nonumber\\
&&+\frac{1}{3}\mathring{T}_{[\nu}t_{2}^{\lambda}\,_{\mu\rho]}+\frac{1}{12}\varepsilon_{\sigma\mu\nu\rho}W^{\lambda}S^{\sigma}+\frac{1}{12}\varepsilon^{\lambda}\,_{\sigma[\mu\nu}W_{\rho]}S^{\sigma}+\frac{1}{2}W_{[\rho}t_{2}^{\lambda}\,_{\mu\nu]}\,.
\end{eqnarray}

In this case, the vector part of the torsion tensor is completely described by the following components
\begin{eqnarray}
\mathring{T}_{r}&=&-\,\frac{\kappa_{d}\Psi_{1}-6rg\Psi_{1}\Psi_{2}-r^{2}g\Psi_{1}\Psi'_{2}-r^{2}g\Psi_{2}\Psi'_{1}-2r^{2}\Psi_{1}\Psi_{2}g'}{2r\Psi_{1}\Psi_{2}}\,,\\
\mathring{T}_{t}&=&-\,\mathring{T}_{r}\sqrt{\Psi_{1}\Psi_{2}}\,,
\end{eqnarray}
whereas the tensor part can be split as $t^{\lambda}\,_{\mu\nu}=\mathring{t}_{1}^{\lambda}\,_{\mu\nu}+t_{2}^{\lambda}\,_{\mu\nu}$, with $t_{2}^{\lambda}\,_{\mu\nu}=\mathring{t}_{2}^{\lambda}\,_{\mu\nu}+\bar{t}^{\lambda}\,_{\mu\nu}$. Hence, this part has at least the following non-propagating components
\begin{eqnarray}
\mathring{t}_{1}^{\,t}\,_{tr}&=&\frac{\kappa_{d}\Psi_{1}-r^{2}g\Psi_{1}\Psi'_{2}-r^{2}g\Psi_{2}\Psi'_{1}-2r^{2}\Psi_{1}\Psi_{2}g'}{3r\Psi_{1}\Psi_{2}}\,,\\
\mathring{t}_{1}^{\,r}\,_{tr}&=&-\,2\mathring{t}_{1}^{\,\theta_{1}}\,_{t\theta_{1}}=-\,2\mathring{t}_{1}^{\,\theta_{2}}\,_{t\theta_{2}}=2\mathring{t}_{1}^{\,\theta_{1}}\,_{r\theta_{1}}\sqrt{\Psi_1\Psi_2}=2\mathring{t}_{1}^{\,\theta_{2}}\,_{r\theta_{2}}\sqrt{\Psi_1\Psi_2}=\mathring{t}_{1}^{\,t}\,_{tr}\sqrt{\Psi_1\Psi_2}\,.
\end{eqnarray}

On the other hand, the second contribution of the antisymmetrized part of the curvature tensor constitutes a general piece associated with the following axial components
\begin{eqnarray}\label{generalsolution2a}
S_{r}&=&-\,\frac{2}{r^{2}\sqrt{\Psi_{2}}}\left(k\sqrt{\Psi_{1}}+2r^{2}h\sqrt{\Psi_{2}}\right)\,,\\
S_{t}&=&-\,S_{r}\sqrt{\Psi_{1}\Psi_{2}}\,,
\end{eqnarray}
as well as with the following tensor ones,
\begin{eqnarray}
t_{2}^{\,\theta_{2}}\,_{t\theta_{1}}&=&\frac{\sqrt{\Psi_{1}}}{3r^{2}\sin\theta_{1}}\left(k\sqrt{\Psi_{1}}-r^{2}h\sqrt{\Psi_{2}}\right)\,,\label{generalsolution2b}\\
t_{2}^{\,\theta_{1}}\,_{t\theta_{2}}&=&-\,t_{2}^{\,\theta_{2}}\,_{t\theta_{1}}\sin^{2}\theta_{1}=-\,t_{2}^{\,\theta_{1}}\,_{r\theta_{2}}\sqrt{\Psi_{1}\Psi_{2}}=t_{2}^{\,\theta_{2}}\,_{r\theta_{1}}\sqrt{\Psi_{1}\Psi_{2}}\sin^{2}\theta_{1}\,,\label{generalsolution2c}\\
t_{2}^{\,r}\,_{\theta_{1}\theta_{2}}&=&2r^{2}t_{2}^{\,\theta_{1}}\,_{r\theta_{2}}\Psi_{2}=t_{2}^{\,t}\,_{\theta_{1}\theta_{2}}\sqrt{\Psi_{1}\Psi_{2}}\,.\label{generalsolution2d}
\end{eqnarray}

All these parts and the torsion tensor itself satisfy the following orthogonal properties in the present case:
\begin{eqnarray}
\delta^{\lambda}\,_{[\rho}\mathring{t}_{1}^{\sigma}\,_{\mu\nu]}\mathring{T}_{\sigma}&=&\delta^{\lambda}\,_{[\rho}t_{2}^{\sigma}\,_{\mu\nu]}\mathring{T}_{\sigma}=\mathring{T}_{[\mu}\mathring{t}_{1}^{\lambda}\,_{\rho\nu]}=\mathring{T}_{[\mu}\mathring{t}_{1}{}_{\rho\nu]}\,^{\lambda}=0\,,
\\
\varepsilon_{\lambda\rho\mu\nu}\mathring{T}^{\mu}S^{\nu}&=&\varepsilon_{\sigma\omega[\mu\nu}\mathring{t}_{1}{}_{\rho]}{}^{\sigma\lambda}S^{\omega}=\varepsilon_{\sigma\omega[\mu\nu}t_{2}{}_{\rho]}{}^{\sigma\lambda}S^{\omega}=0\,,
\\
\mathring{t}_{1}^{\sigma}\,_{[\mu\nu}\mathring{t}_{1}{}_{\rho]\sigma}{}^{\lambda}&=&t_{2}^{\sigma}\,_{[\mu\nu}t_{2}{}_{\rho]\sigma}{}^{\lambda}=\mathring{t}_{1}^{\sigma}{}_{[\mu\nu}\mathring{t}_{1}^{\lambda}{}_{\rho]\sigma}=t_{2}^{\sigma}{}_{[\mu\nu}t_{2}^{\lambda}{}_{\rho]\sigma}=0\,,
\\
\delta^{\lambda}\,_{[\rho}T^{\sigma}\,_{\mu\nu]}W_{\sigma}&=&\delta^{\lambda}\,_{[\rho}W_{\mu}\mathring{T}_{\nu]}=W_{[\rho}\mathring{t}_{1}^{\lambda}\,_{\mu\nu]}=0\,,
\\
\varepsilon_{\sigma\rho\mu\nu}\mathring{T}^{\lambda}S^{\sigma}&=&-\,3\,\varepsilon^{\lambda}\,_{\sigma[\rho\mu}\mathring{T}_{\nu]}S^{\sigma}\,,
\\
T_{\sigma}\,^{\lambda}\,_{[\rho}T^{\sigma}\,_{\mu\nu]}&=&0
\,.\label{fundamentalproperties}
\end{eqnarray}

Accordingly, the vector mode of torsion is fully constrained by the Eq.~(\ref{curvaturedecomposition2}) and the dynamical character of torsion is governed by the corresponding parts $\bar{S}_{\mu}$ and $\bar{t}^{\lambda}\,_{\mu\nu}$ contained in $\tilde{R}_{2}^{\lambda}\,_{[\mu \nu \rho]}$ (i.e. $\bar{T}_{\mu}=0$). In any case, the above relations do not set the values of the remaining parts $\mathring{S}_{\mu}$ and $\mathring{t}_{2}^{\lambda}\,_{\mu\nu}$, so they must be certainly obtained by the field equations (\ref{field_eq1}) and (\ref{field_eq2}). This means that in general, it is also possible to find nontrivial solutions related to a non-propagating component $\mathring{R}_{2}^{\lambda}\,_{[\mu\nu\rho]}$ depending on these parts. On the other hand, the absence of a propagating vector mode for torsion in the dynamical regime involves that the weak-field limit (\ref{irreduciblelimit}) in that case is reduced to the following constraint involving the tensor mode,
\begin{equation}\label{dynamicaltensorlimit}
\nabla_{\mu}\nabla_{\lambda}t_{2}^{\lambda\mu\nu}=0\,,
\end{equation}
which is satisfied by the respective components of the Eqs.~(\ref{generalsolution2b})-(\ref{generalsolution2d}), as expected. Furthermore, the total antisymmetric part of the Ricci tensor also fulfills
\begin{equation}
\nabla_{\mu}\tilde{R}^{[\mu\nu]}=0\,.
\end{equation}
This means that the dynamical contribution provided by this part of the Ricci tensor can be equivalently described by the weak-field limit (\ref{dynamicaltensorlimit}) in the present case.

\subsection{Exact black hole solutions}

We have seen how the weak-field limit described by Eqs.~(\ref{weaktorsion}) and (\ref{weaknonmetricity}), as well as the consistency constraints (\ref{rel1}) and (\ref{rel2}), allow the dynamical contributions of torsion and nonmetricity to be decomposed in a regular way. In addition, in the absence of matter fields, we also know that the static and spherically symmetric vacuum configuration with dynamical torsion related to the gravitational action (\ref{LagrangianIrreducible}) requires the condition $c_{2}=2c_{1}=-\,d_{1}/2$ under these consistency constraints \cite{Cembranos:2016gdt,Cembranos:2017pcs}. In fact, this particular combination of the Lagrangian coefficients allows the dynamical kinetic terms of that case to be written in the following simplified regular form,
\begin{equation}
S_{\rm kin}=\; \frac{d_{1}}{32\pi}\int d^4x \sqrt{-g} \left(\frac{1}{9}\partial_{\mu}\bar{S}_{\nu}\left(\partial^{\mu}\bar{S}^{\nu}-\partial^{\nu}\bar{S}^{\mu}\right)+\partial_{\lambda}\bar{t}^{\lambda}\,_{\mu\nu}\partial_{\rho}\bar{t}^{\rho\mu\nu}+3\,\partial_{\rho}\bar{t}_{\lambda\mu\nu}\left(\partial^{\rho}\bar{t}^{\mu\nu\lambda}-\partial^{\mu}\bar{t}^{\lambda\nu\rho}\right)\right)\,.
\end{equation}

It is worthwhile to emphasize the dynamical role of the torsion and nonmetricity fields, which indeed modify the structure of the background space-time and shows nontrivial couplings with the Riemannian sector even in the weak-field regime given by the Expressions (\ref{weaktorsion}) and (\ref{weaknonmetricity}). This fact points out the relevance of such a background space-time when studying the stability of post-Riemannian theories since they can show certain instabilities emerging from this interaction, which in any case can be stabilised by the introduction into the gravitational action of higher order corrections depending on mixing terms \cite{Jimenez:2019qjc}.

By taking into account the compatibility of such a vacuum structure with the energy-momentum tensor provided by Coulomb electric and magnetic charges, the condition $\Psi_{1}(r)=\Psi_{2}(r) \equiv \Psi(r)$ must also be fulfilled in order to satisfy the Maxwell’s equations of the electromagnetic field. This equivalence allows the integration of the Eq.~(\ref{rel3}), which gives rise to the Coulomb potential of the Tresguerres solution for the dilation charge \cite{Tresguerres:1995js}:
\begin{equation}\
w_{1}(r)=\frac{\kappa_{d}}{r}\,.
\end{equation}

Furthermore, the symmetric component in the indices $\lambda$ and $\mu$ of the field equation (\ref{field_eq2}) is also trivially reduced to the constraint $e_{2}=-\,d_{1}/2$, which places the constants $d_{1}$, $e_{1}$ and $1-2a_{2}$ as the three independent Lagrangian coefficients of the model. The antisymmetric components of (\ref{field_eq2}) with $\lambda = r, \mu = \nu = t$ and $\lambda = t, \mu = \nu = \theta_{1}$, as well as the antisymmetric one with $\mu = t$ and $\nu = r$ of (\ref{field_eq1}) provide then the following system of independent equations for the functions $h(r)$ and $k(r)$:
\begin{eqnarray}
0&=&k\left(r\Psi'h+r\Psi h' +3h \Psi\right)\,,\\
0&=&2r^4 h \Psi h'+2r^4h^2 \Psi'+3r^2k\Psi h'+5r^2 k h \Psi'+r k h \Psi-2k^2 \Psi'-2k\Psi k'+2r^2 h \Psi k'+2r^3 h^2 \Psi\,,\\
0&=&2r^3h^2\Psi^2-k^2\Psi\Psi'-r^2k h \Psi \Psi'-r^3 k h \Psi'^{2}-r^3 k \Psi h' \Psi'-r^2 h \Psi^2 k'+2r k h \Psi^2-k \Psi^2 k'+2 r^4 h \Psi^2 h'
\nonumber\\
&&+2 r^4 h^2 \Psi \Psi'-r^3\Psi^2 k'h'-r^3 h \Psi\Psi' k'\,.
\end{eqnarray}

The above system of equations has as unique solutions either $k(r)=0, \, h(r)=-\,\kappa_{s}/(r\Psi(r))$ or $h(r)=0,\, k(r)=\beta/\Psi(r)$, where $\kappa_{s}$ and $\beta$ are integration constants. Nevertheless, the latter gives rise to a nontrivial solution of the remaining field equations characterized by a vanishing component $\tilde{R}_{2}^{\lambda}\,_{[\mu\nu\rho]}$ of the field strength tensor (i.e. $\bar{S}_{\mu}=\bar{t}^{\lambda}\,_{\mu\nu}=0$), which means that $k(r)=0$ and $h(r)=-\,\kappa_{s}/(r\Psi(r))$ in the dynamical case, in such a way that $\kappa_{s}$ represents the spin charge of the gravitational system.

Therefore, the component with $\lambda = \theta_{1}, \mu = \theta_{2}$ and $\nu = t$  of the Eq.~(\ref{field_eq2}) trivially allows us to isolate the last function of the torsion tensor as follows
\begin{equation}
g(r)=-\,\frac{1}{2r}-\frac{wr}{2\Psi(r)}-\frac{\kappa_{d}}{2r\Psi(r)}\,,
\end{equation}
with $w=\left(1-2a_{2}\right)/d_{1}$. Finally, Eq.~(\ref{field_eq1}) is reduced to the following independent equation depending on the metric function
\begin{equation}
r^2-r^2\Psi(r)-r^3\Psi'(r) = d_{1}\kappa_{s}^{2}-4e_{1}\kappa_{d}^{2}\,,
\end{equation}
which has as solution, the Reissner-Nordstr\"{o}m metric
\begin{eqnarray}
\Psi(r)&=&1-\frac{2m}{r}+\frac{d_{1}\kappa^{2}_{s}-4e_{1}\kappa^{2}_{d}}{r^2}\,.
\end{eqnarray}

In summary, the new exact black hole solution related to the gravitational action (\ref{LagrangianIrreducible}) in the absence of matter fields is described by the following expressions:
\begin{eqnarray}\label{solution}
a(r)&=&\frac{\Psi'(r)}{2\Psi(r)}+\frac{wr}{\Psi(r)}+\frac{\kappa_{d}}{2r\Psi(r)}\,,\quad
b(r)=\frac{\Psi'(r)}{2}+wr+\frac{\kappa_{d}}{2r}\,,\;\;\;\;\;\;\;\;\;\;\;\;\;\;\;\;
\nonumber\\
c(r)&=&\frac{\Psi(r)}{2r}+\frac{wr}{2}+\frac{\kappa_{d}}{2r}\,,\quad
g(r)=-\,\frac{1}{2r}-\frac{wr}{2\Psi(r)}-\frac{\kappa_{d}}{2r\Psi(r)}\,, \nonumber\\
d(r)&=&\frac{\kappa_{s}}{r}\,,\quad
h(r)=-\,\frac{\kappa_{s}}{r\Psi(r)}\,,\quad
k(r)=l(r)=0\,,
\\
w_{1}(r)&=&\frac{\kappa_{d}}{r}\,,\quad
w_{2}(r)=-\,\frac{\kappa_{d}}{r\Psi(r)}\,,
\nonumber\\
\Psi(r)&=&1-\frac{2m}{r}+\frac{d_{1}\kappa^{2}_{s}-4e_{1}\kappa^{2}_{d}}{r^2}\,,\nonumber
\end{eqnarray}
with
\begin{equation}\label{constants}
c_2=2c_1=e_2=-\,\frac{d_1}{2} \,, \quad w=\frac{\left(1-2a_{2}\right)}{d_{1}}\,.
\end{equation}

As can be seen, the solution describes a Reissner-Nordstr\"{o}m type of geometry provided by both spin and dilation charges, in virtue of the fundamental relation of torsion and nonmetricity with their spinning and dilational sources. Indeed, the dynamical components of torsion and nonmetricity act as Coulomb-like potentials depending on $\kappa_{s}$ and $\kappa_{d}$, which are consistently induced in the gravitational scheme by the following field strength tensors of the interaction
\begin{eqnarray}
\tilde{R}^{\sigma}\,_{\sigma\mu\nu}&=&4\nabla_{[\nu}W_{\mu]}\,,\\
\tilde{R}^{\lambda}\,_{[\mu\nu\rho]}&=&\frac{1}{4}\tilde{R}^{\sigma}\,_{\sigma[\mu\nu}\delta_{\rho]}\,^{\lambda}+\bar{R}^{\lambda}\,_{[\mu\nu\rho]}\,,\\
\tilde{R}_{[\mu\nu]}&=&\frac{1}{4}\tilde{R}^{\sigma}\,_{\sigma\mu\nu}+\bar{R}_{[\mu\nu]}\,,\\
\bar{R}^{\lambda}\,_{[\mu \nu \rho]}&=&\frac{1}{6}\varepsilon^{\lambda}\,_{\sigma[\rho\nu}\nabla_{\mu]}\bar{S}^{\sigma}+\nabla_{[\mu}\bar{t}^{\lambda}\,_{\rho\nu]}-\frac{1}{18}\varepsilon_{\sigma\mu\nu\rho}\mathring{T}^{\lambda}\bar{S}^{\sigma}+\frac{1}{4}\varepsilon^{\lambda}\,_{\omega\sigma[\rho}\mathring{t}_{1}^{\sigma}\,_{\mu\nu]}\bar{S}^{\omega}+\frac{1}{12}\varepsilon_{\sigma\mu\nu\rho}W^{\lambda}\bar{S}^{\sigma}\,,\\
\bar{R}_{[\mu\nu]}&=&\frac{1}{12}\varepsilon^{\lambda}\,_{\sigma\mu\nu}\nabla_{\lambda}\bar{S}^{\sigma}+\frac{1}{2}\nabla_{\lambda}\bar{t}^{\lambda}\,_{\mu\nu}\,,
\end{eqnarray}
where the parts $\bar{S}_{\mu}$ and $\bar{t}^{\lambda}\,_{\mu\nu}$ coincide with the axial and tensor components of Eqs~(\ref{generalsolution2a})-(\ref{generalsolution2d}) particularized for the solution (i.e. $\mathring{S}_{\mu}=\mathring{t}_{2}^{\lambda}\,_{\mu\nu}=0$ for the solution). Thus, the antisymmetrized curvature tensor is expressed as the sum of a degenerate homothetic component related to nonmetricity and a dynamical contribution governed by the previous parts of torsion, as expected. Furthermore, it satisfies the following Proca-like equation
\begin{equation}\label{procalike}
\nabla_{\lambda}\tilde{R}^{\lambda}\,_{[\rho \mu \nu]}-w\,T_{[\rho\mu\nu]}=0\,.
\end{equation}

The existence of a dynamical axial mode of torsion in the solution is especially relevant, since this is the unique component implicated in the interaction with Dirac fields, according to the minimal coupling principle \cite{Shapiro:2001rz,Cembranos:2018ipn}. Furthermore, the inclusion of a nontrivial Weyl vector involves an additional isotropic change in the volume of dilational test bodies, which enables an extended description of the dynamical effects caused by these post-Riemannian quantities in terms of the corresponding multipole moments of such deformable test bodies with microstructure \cite{Puetzfeld:2014qba,Obukhov:2015eqa}. In any case, as is shown, the dynamical character of torsion and nonmetricity is induced on the metric and coframe forms by the field equations and thereby they can also operate on the geodesic motion of ordinary matter via the Levi-Civita connection. The magnitude of the deviations derived from these quantities depends on the value of both the spin and dilation charges of the source and of the coupling constants that determine the fundamental strength of the interaction. In any case, significant effects are contemplated only around extreme gravitational systems, such as neutron stars with intense magnetic fields and sufficiently oriented elementary spins or black holes endowed with spin and dilation charges, in virtue of the purely quantum nature of the intrinsic angular momentum of matter and the expected vanishing of dilation currents in the rest of ordinary matter sources. In this sense, the performance of new phenomenological analyses following these lines is especially desirable for the design of future tests of gravity beyond the Einstein's framework \cite{Hammond:2002rm,Rubiera-Garcia:2020gcl}.

On the other hand, as mentioned previously, the fulfillment of the Maxwell’s equations of the electromagnetic field trivially leads to a simple generalization of the solution when considering both Coulomb electric and magnetic charges $q_{e}$ and $q_{m}$, but also a nonvanishing cosmological constant $\Lambda$ in the gravitational action, by replacing the metric function as follows
\begin{equation}
\Psi(r)=1-\frac{2m}{r}+\frac{d_{1}\kappa^{2}_{s}-4e_{1}\kappa^{2}_{d}+q^{2}_{e}+q^{2}_{m}}{r^2}+\frac{\Lambda}{3}r^{2}\,.
\end{equation}
This is the most important result of this work since we have found a new exact black hole solution in this metric-affine theory of gravity with independent spin and dilation charges.

\section{Conclusions}\label{sec:conclusions}
In the present work, we propose a gravitational model based on Weyl-Cartan geometry containing the minimal deviations from GR which enable an independent behaviour for the torsion and nonmetricity tensors to be displayed in a consistent way. Accordingly, the model allows the existence of a black hole solution provided by Coulomb-like potentials for the dynamical components of torsion and nonmetricity, which constitutes a fully relativistic configuration with independent dynamical torsion and nonmetricity fields (i.e. beyond the well-known black hole solutions provided by the triplet ansatz and other similar constraints in the framework of MAG). In this sense, the present solution shows similarities among the torsion and nonmetricity parts of the previous solutions as well as with the electromagnetic field, even though they constitute independent quantities. The presence of a nontrivial axial mode that propagates in a reasonable way at large distances is especially relevant since this is the unique component of the torsion tensor implicated in the interaction with Dirac fields, according to the minimal coupling principle, and enables the development of future phenomenological tests of strong gravity beyond the Einstein's framework. Furthermore, the combined dynamical effects induced by torsion and nonmetricity in the metric tensor itself are expected to play a crucial role in the achievement of a full post-Riemannian description of the quasinormal modes of gravitational waves emitted by the mergers of compact objects, such as neutron stars or black holes with torsion and nonmetricity, in virtue of the intrinsic relation between the underlying axial and polar perturbations and the structure of the background space-time \cite{Abbott:2016blz,TheLIGOScientific:2017qsa,Abbott:2020uma,Abbott:2020tfl,Kokkotas:1999bd}. 

It is also worthwhile to stress some other natural generalizations of these results, such as the search of the rotating counterpart of the present black hole solution in an axisymmetric space-time and/or the inclusion of shear charges into the dynamical scheme. In addition, a large class of studies on the possible implications of the torsion and nonmetricity fields at cosmological scales have also been carried out by many authors with promising results, such as the avoidance of initial singularities, the generation of inflationary and late-time acceleration phases in the evolution of the universe, or new dark matter gravitational interactions, among others \cite{Puetzfeld:2004yg,Minkevich:2009df,Chen:2009at,Poplawski:2011jz,Magueijo:2012ug,Lu:2016bcx,BeltranJimenez:2017vop,Iosifidis:2020gth}. The search of new cosmological configurations defined in the metric-affine geometry of the present model is also relevant to analyse the behaviour of the torsion and nonmetricity fields, especially in early stages of the universe. Further research following these lines of study will be addressed in future works.

\bigskip
\bigskip
\noindent
\section*{Acknowledgements}
The authors would like to thank Yuri N. Obukhov for helpful discussions. This work was supported by the European Regional Development Fund and the programme Mobilitas Pluss (Grants No. MOBJD423 and MOBJD541).
\newpage
\appendix
\section{Table with the list of conventions}

\begin{table}[h]
\begin{center}
\begin{tabular}{|c|l|}
\hline
$\tilde{R}^{\lambda}{}_{\rho\mu\nu}$ &  Curvature tensor with general affine connection \\  [0.7ex] 
$R^{\lambda}{}_{\rho\mu\nu}$& Curvature tensor with Levi-Civita connection \\ [0.7ex] 
$\bar{R}^{\lambda}{}_{\rho\mu\nu}$& Propagating mode of the curvature tensor \\[0.7ex] 
$\mathring{R}^{\lambda}{}_{\rho\mu\nu}$& Non-propagating mode of the curvaure tensor \\
[0.7ex] $\tilde{\Gamma}^{\lambda}{}_{\mu\nu}$& General affine connection \\   [0.7ex]
${\Gamma}^{\lambda}{}_{\mu\nu}$& Levi-Civita connection \\   [0.7ex]
$\tilde{\nabla}_\mu$& Covariant derivative with respect to the general affine connection  \\   [0.7ex]
$\nabla_\mu$& Covariant derivative with respect to the Levi-Civita connection \\   [0.7ex]
$\omega^{a}\,_{b\nu}$ & Anholonomic general affine connection \\[0.7ex]
$g_{ab}$ & Anholonomic metric tensor \\[0.7ex]
$e^a{}_\mu$ & Tetrad field \\[0.7ex]
$T^{\lambda}{}_{\mu\nu}\equiv  2\tilde{\Gamma}^{\lambda}\,_{[\mu\nu]}$& Torsion tensor \\    [0.7ex]
$T_{\mu}\equiv T^{\lambda}{}_{\mu\lambda}$& Vector part of torsion \\    [0.7ex]
$S_{\mu}\equiv \epsilon_{\mu \lambda \rho \nu}T^{\lambda\rho\nu}$& Axial part of torsion \\    [0.7ex]
$t^{\lambda}{}_{\mu\nu}$& Tensor part of torsion \\    [0.7ex]
$\bar{T}_{\mu}, \ \bar{S}_\mu ,\ \bar{t}^{\lambda}{}_{\mu\nu}$& Dynamical vector, axial and tensor parts of torsion \\    [0.7ex]
$\mathring{T}_{\mu}, \ \mathring{S}_\mu ,\ \mathring{t}^{\lambda}{}_{\mu\nu}$& Non-propagating vector, axial and tensor parts of torsion \\    [0.7ex]
$Q_{\lambda\mu\nu}\equiv \tilde{\nabla}_\lambda g_{\mu\nu}$& Nonmetricity tensor \\    [0.7ex]
$W_{\mu} \equiv \frac{1}{4}Q_{\mu\lambda}\,^{\lambda}$& Weyl vector \\    [0.7ex]
$\tilde{R}_{\mu\nu}\equiv \tilde{R}^{\lambda}{}_{\mu\lambda\nu}$& Ricci tensor with general affine connection \\    [0.7ex]
$R_{\mu\nu}\equiv R^{\lambda}{}_{\mu\lambda\nu}$& Ricci tensor with Levi-Civita connection \\    [0.7ex]
$\hat{R}_{\mu\nu}\equiv \tilde{R}_\mu{}^{\lambda}{}_{\nu\lambda}$& Co-Ricci tensor with general affine connection \\    [0.7ex]
$\tilde{R}^{\lambda}{}_{\lambda\mu\nu}$& Homothetic curvature tensor with general affine connection \\    [0.7ex]
$\tilde{R}\equiv \tilde{R}^{\lambda\rho}{}_{\lambda\rho}$& Scalar curvature with general affine connection \\    [0.7ex]
$R\equiv R^{\lambda\rho}{}_{\lambda\rho}$& Scalar curvature with Levi-Civita connection \\    [0.7ex]
$\tilde{G}_{\mu\nu}\equiv \tilde{R}_{\mu\nu}-\frac{1}{2}\tilde{R}g_{\mu\nu}$& Einstein tensor with general affine connection \\    [0.7ex]
$G_{\mu\nu}\equiv R_{\mu\nu}-\frac{1}{2}Rg_{\mu\nu}$& Einstein tensor with Levi-Civita connection \\    [0.7ex]
$A_{(\mu\nu}{}^{\lambda}{}_{\sigma)}$& Symmetrization of a tensor $A_{\mu\nu}{}^{\lambda}{}_{\sigma}$ with respect to the covariant indices \\    [0.7ex]
$A_{[\mu\nu}{}^{\lambda}{}_{\sigma]}$& Antisymmetrization of a tensor $A_{\mu\nu}{}^{\lambda}{}_{\sigma}$ with respect to the covariant indices \\  [0.7ex]
\hline
\end{tabular}
\caption{List of notational conventions.}
\label{tab:notation}
\end{center}
\end{table}

\bibliographystyle{utphys}
\bibliography{references}

\end{document}